\documentclass[12pt]{iopart}
\usepackage{graphicx}
\usepackage{enumerate}

\begin{document}

\title[Cohesive motion in one-dimensional flocking]{Cohesive motion in one-dimensional flocking}

\author{V Dossetti}

\address{$^1$ Instituto de F\'{\i}sica, Benem\'erita Universidad Aut\'onoma de Puebla, Apartado Postal J-48, Puebla 72570, Mexico}
\address{$^2$ Consortium of the Americas for Interdisciplinary Science and Department of Physics and Astronomy, University of New Mexico, Albuquerque, NM 87131, USA}
\ead{dossetti@ifuap.buap.mx}

\begin{abstract}
A one-dimensional rule-based model for flocking, that combines velocity alignment and long-range centering interactions, is presented and studied. The induced cohesion in the collective motion of the self-propelled agents leads to a unique group behaviour that contrasts with previous studies. Our results show that the largest cluster of particles, in the condensed states, develops a mean velocity slower than the preferred one in the absence of noise. For strong noise, the system also develops a non-vanishing mean velocity, alternating its direction of motion stochastically. This allows us to address the directional switching phenomenon. The effects of different sources of stochasticity on the system are also discussed.
\end{abstract}

\pacs{05.40.-a, 05.70.Fh, 64.60.Cn, 87.10.-e}


\section{Introduction}
The phenomenon of flocking (also known as swarming) refers to the emergent collective motion regularly seen in groups of animals such as bird flocks, fish schools, bison or sheep herds, insect or bacteria swarms, etc. \cite{Lor00, Bon98, Par99, Cou05, Buh06, Yat09, Wu00} In simple terms, the formation of these sort of groups in nature, implies some kind of condensation in position and/or velocity spaces. The mechanisms or behaviours followed by the individuals that compose the group, that lead to this type of condensation, have been classified by Reynolds \cite{Rey87} as: the desire to match the velocity of flockmates (alignment), the desire to stay close to flockmates (centering), and the desire to avoid collisions (separation). These definitions have been widely adopted by researchers from a variety of disciplines.

In the realm of the physical sciences, the study of flocking has greatly benefited from the insight of statistical mechanics to describe dynamical phase transitions and cooperative phenomena \cite{Ton05}. The interest in the study of this kind of systems comes from the fact that they represent prototypical out-of-equilibrium systems. The formulations to study the flocking phenomenon can be succinctly classified in rule-based \cite{Vic95, Gre03, Nag07, Per08, Bus97, Ald03, Olo99, Ray06, Bod10}, Lagrangian (trajectory-based) \cite{Mik99, Erd05, Shi96, Lev00, Dor06, Dos09, Str08, Iwa10} and Eulerian (continuum) models \cite{Top04, Ber06, Mog96}. Regarding the dimensionality of the models developed, most of them have been defined in dimensions higher than one \cite{Vic95, Gre03, Nag07, Per08, Bus97, Ald03, Erd05, Shi96, Lev00, Dor06, Dos09, Str08, Top04, Ber06, Mog96, Gon08}, since the velocity of the self-propelled particles (SPP) in these models can have continuous values. In contrast, only a few one-dimensional (1D) models have been studied  \cite{Buh06, Olo99, Ray06, Bod10, Mik99, Iwa10, Czi99}. For these, the direction of the particles velocity can only take discrete values, in fact, only two. Nonetheless, recent experimental and theoretical studies by Buhl \etal \cite{Buh06} and Yates \etal \cite{Yat09} have proven the usefulness of 1D models in elucidating the \emph{directional switching} phenomenon. This phenomenon regards a very interesting aspect of the movement of many animal groups, when they suddenly switch their direction of motion in a coherent form. This seems to be an intrinsic property of the collective motion of animals as it has been observed, e.g., in groups of locusts and starlings (see \cite{Yat09} for more details).

Among the one-dimensional models, one can find of the Lagrangian kind like the one by Mikhailov  and Zanette \cite{Mik99}, that describes a population of self-propelled particles (SPP) with attractive long-range interactions, i.e., includes centering as the only ingredient from the Reynolds considerations. Here, the particles try to attain one of two preferred velocities ($v_0$ or -$v_0$) through a non-linear friction term cubic in nature. The system presents multistable dynamics and can be found either in coherent traveling motion (\emph{ordered state}) where all the particles move together in a condensed group, developing a non-zero mean velocity with the same direction, or in an incoherent oscillatory state (\emph{disordered state}) where the particles perform random-phase oscillations around a fixed position with zero mean velocity. By increasing the additive noise intensity, the system can be driven from the coherent traveling state to the oscillatory state through a discontinuous phase transition. However, for noise intensities below the critical point, the two states are accessible depending on the initial conditions since the system shows a strong hysteresis.

On the other hand, one can find rule-based models like the 1D version of the famous Vicsek \etal model \cite{Vic95}, by Czir\'ok \etal \cite{Czi99}, that describes a population of SPP that interact only through condensation in velocity space (alignment according to Reynolds). Via short-range interactions and a majority rule, the particles try to match the velocity of their nearest flockmates. The system exhibits spontaneous symmetry breaking and self-organization depending on the intensity of the also additive noise, leading to a continuous kinetic phase transition from a homogenous to a condensed phase. In contrast with the 2D Vicsek model \cite{Vic95} where the speed of the particles is fixed, i.e., $|v_0|=\hbox{constant}$, in the 1D version \cite{Czi99}, at each time step, particles try to acquire a velocity close to the two prescribed ones ($v_0$ or $-v_0$) through an antisymmetric function $G(u)$ that depends linearly on the local mean velocity $\left<u(t)\right>_i$ of the neighbours of a given particle $i$. This function can also be written in a continuous form with no change in the scaling properties of the system as claimed in \cite{Czi99}.

Let us also consider the lattice model by O'Loan and Evans \cite {Olo99} where the velocity of the particles can take three discrete values ($-1, 0 ,1$). Here, through an alignment interaction in the form of a majority rule, that is applied asynchronously with probabilistic updates (particles are selected and updated randomly in a Monte Carlo fashion), the system shows a continuous phase transition from a homogeneous to a condensed phase similar to that of the previous model. However, and in contrast with the previous model, the condensed phase is not symmetry-broken but alternating. We must advert the two sources of noise in this model --- coming from the way particles are updated and how the alignment rule is applied --- since no additive noise is present. The later generalization of this model by Raymond and Evans \cite{Ray06} includes all three Reynolds' flocking behaviours that translates in richer dynamics: the system also shows an homogeneous flock with homogeneous density and a fixed global non-zero velocity, and dipole states. However, the different interactions present in these models are local, i.e., short in range regarding the spatial coordinate.

Finally, it is worth to point out the variations of the two previous rule-based models introduced by Buhl \etal \cite{Buh06}, Yates \etal \cite{Yat09} and Bode \etal \cite{Bod10}, that also show an alternating condensed phase that resembles the directional switching phenomenon mentioned before. In the first, Buhl \etal consider a generalization of the Czir\'ok \etal model, that weights the local mean direction of motion of the neighbours of a given particle with that of the particle itself in a deterministic way. In the second, Yates \etal adapted the Buhl \etal model with an additive noise whose intensity is weighted with a non-trivial function of the local mean velocity: the individuals increase the randomness of their motion when the local alignment is low. In the third, Bode \etal mix features from the Czir\'ok \etal,  O'Loan and Evans, and Raymond and Evans models, concentrating on how noise enters in the system. There, they have identified three different approaches (that can be generalized to higher dimensions) to include stochastic errors in 1D SSP models and classified them as: 
\begin{enumerate}[($i$)]
\item Adding a random variable to the preferred direction of individuals (i.e., the inclusion of additive noise as in \cite{Mik99, Czi99}).
\item Asynchronous and probabilistic updates (as in \cite{Olo99, Ray06}). 
\item Varying the probability and accuracy with which individuals execute their behavioural rules (as in \cite{Olo99, Ray06} also).
\end{enumerate}
In this way, Bode \etal have included local alignment and centering interactions in their model, that are applied taking inspiration from the approaches ($ii$) and ($iii$). Nonetheless, the interactions considered in all of these models are short-ranged from the point of view of the position of the particles as well.

Our aim in this work is to extend the 1D model of Czir\'ok \etal \cite{Czi99} by providing it with a long-range centering interaction (much in the spirit of Mikhailov and Zanette) along with the already included local alignment interaction and additive noise. We demonstrate a variety of new flocking regimes: some of them resemble those found in the models described in the previous paragraph. We also show that the noise induced phase transition, from the homogeneous to the symmetry-broken condensed phases, changes into what seems a phase transition between two kinds of condensed states: one with broken symmetry and one with an alternating flock that moves with non-zero mean velocity, changing its direction of motion stochastically. The latter seems stable even for strong noise and different system sizes. Additionally, we explore two ways to implement the centering and alignment interactions --- one probabilistic and one deterministic --- that allows us to address the relation of the directional switching phenomenon with different sources of stochasticity in 1D SPP models.

\begin{figure}[t]
\begin{center}
\includegraphics[width=\textwidth]{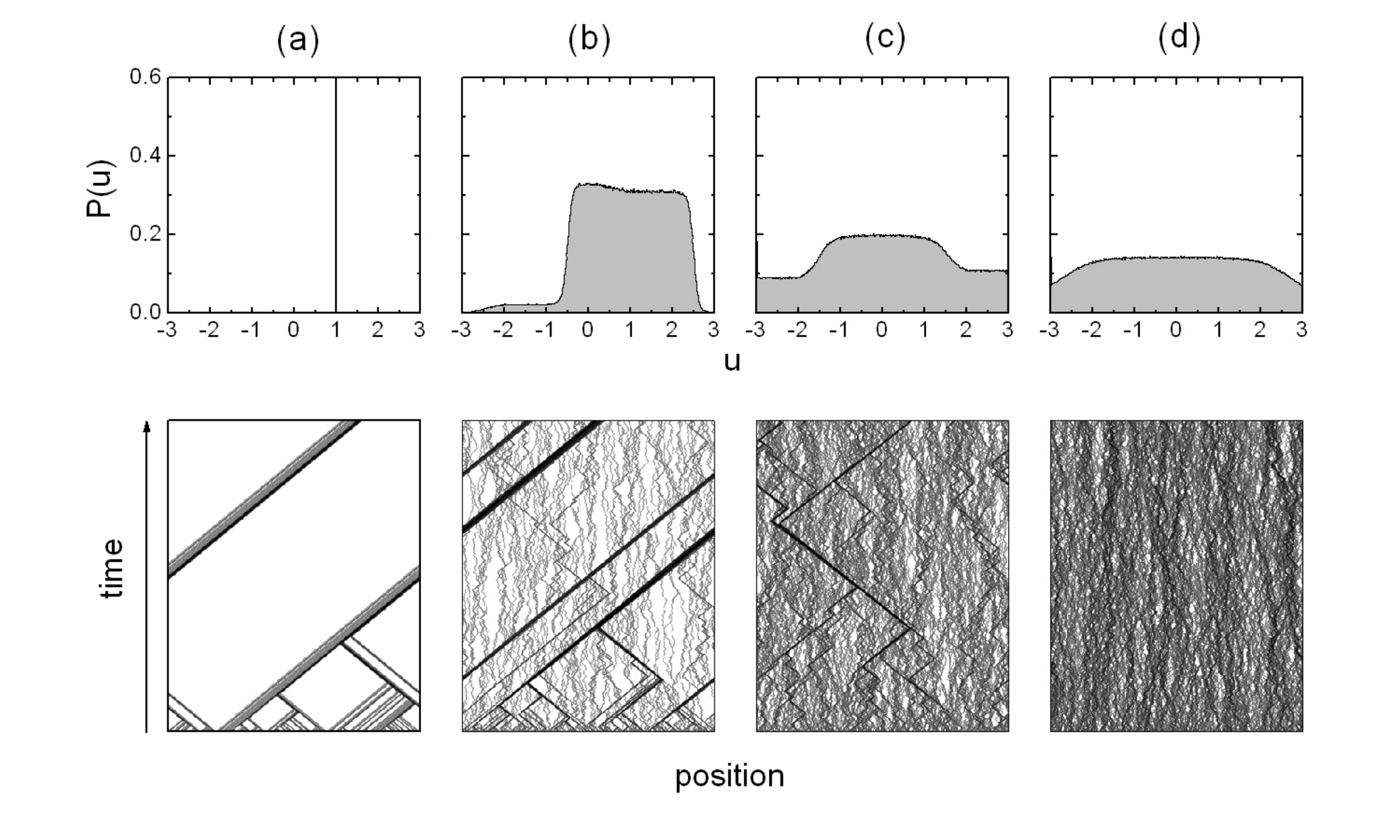} 
\caption{In the lower plots, the dynamics of the 1D Czir\'ok \etal model for $L=1000$, $N=1000$ and $\rho_{\rm a}=1$, for increasing values of $\eta$ (from left to right $\eta=0,3,5,7$, respectively). The darker grey level represents higher particle density and the systems were let to evolve 15000 time steps. The upper plots present the corresponding normalized distribution $P(u)$ of the velocities $u_i$ of the individual particles in the steady state. Notice how the steady state of the system changes from one with broken symmetry to a homogeneous one as the noise amplitude increases.}
\label{fig:v1d-rho1}
\end{center}
\end{figure}

\section{Model}
The 1D model by Czir\'ok \etal \cite{Czi99}, consists of $N$ off-latice SPP along a line of length $L$. The particles are characterized by their coordinate $x_i$ and dimensionless velocity $u_i$ updated as
\begin{equation}
\eqalign{x_i(t+1) = x_i(t) + v_0 u_i(t), \cr
u_i(t+1) = G(\left<u(t)\right>_i) + \xi_i.}
\label{eq:1DVicsek1}
\end{equation}
The local average velocity $\left<u(t)\right>_i$ for the $i$th particle is calculated within an interval $[x_i-\Delta,x_i+\Delta]$, where $\Delta=1$ is fixed, and updated for all particles at every time step. Propulsion and friction forces are incorporated through the antisymmetric function $G$ which sets the velocity in average to a prescribed value $v_0$, in such a way that $G(u)>u$ for $0<u<1$ and $G(u)<u$ for $u>1$. In this case, one cannot apply the constant velocity constraint of \cite{Vic95} since it strongly discretizes the system, leading to a discrete state cellular automaton. It is also important to remark that, similarly with the Raymond and Evans model \cite{Ray06}, function $G(u)$ also works as the interaction term. The noise, $\xi_i$, is distributed uniformly in the interval $[-\eta/2,\eta/2]$ and affects directly the velocity of the particles, while an update time $\Delta t=1$ has already been applied.

In this work, $v_0$ is kept constant $(v_0=0.1)$, and the adjustable control parameters for the model described above are the average density of the particles, $\rho_{\rm a}=N/L$, and the noise amplitude $\eta$. Function $G$ was implemented in \cite{Czi99} in a very simple way as,
\begin{equation}
G(u) = \frac{1}{2}\left[ u + \hbox{sign}(u) \right].
\label{eq:1DVicsek2}
\end{equation}
The results do not depend on the properties of $G(u)$ for $u\approx0$ as stated in \cite{Czi99}. In our case, simulations with continuous choices of $G(u)$ were carried out leading to similar results. Random initial conditions are applied along with periodic boundary conditions.

In the lower plots of figure \ref{fig:v1d-rho1}, we show the time evolution of the density of particles, $\rho(x,t)$, and the normalized distribution $P(u)$ of the velocities $u_i$ of the particles, in the upper plots, for the model given in (\ref{eq:1DVicsek1}) and (\ref{eq:1DVicsek2}) with $L=1000$, $N=1000$ and increasing values of the noise amplitude $\eta(\in[0,7])$. The corresponding average density of particles is $\rho_{\rm a}=1$. Depending on the chosen noise amplitude, $\eta$, the system can be driven through an order-disorder continuous phase transition. For example, for $\eta < \eta_c \approx 5$, the system reaches an ordered state in a short time (see figures \ref{fig:v1d-rho1}(a) and \ref{fig:v1d-rho1}(b)), characterized by a spontaneous symmetry breaking and clustering of the particles. On the contrary, for larger values of $\eta \, (> \eta_{\rm c})$, a disordered phase can be found characterized by a random velocity field: see, e.g., figure \ref{fig:v1d-rho1}(d). As one increases $\eta$, one can appreciate how the distribution $P(u)$ in the steady state, that starts single-peaked and narrow for $\eta=0$ (see upper plots in figure \ref{fig:v1d-rho1}), \emph{flattens} and \emph{broadens} out when going from the ordered to the disordered state (from left to right).

\begin{figure}[tb]
\begin{center}
\includegraphics[width=\textwidth]{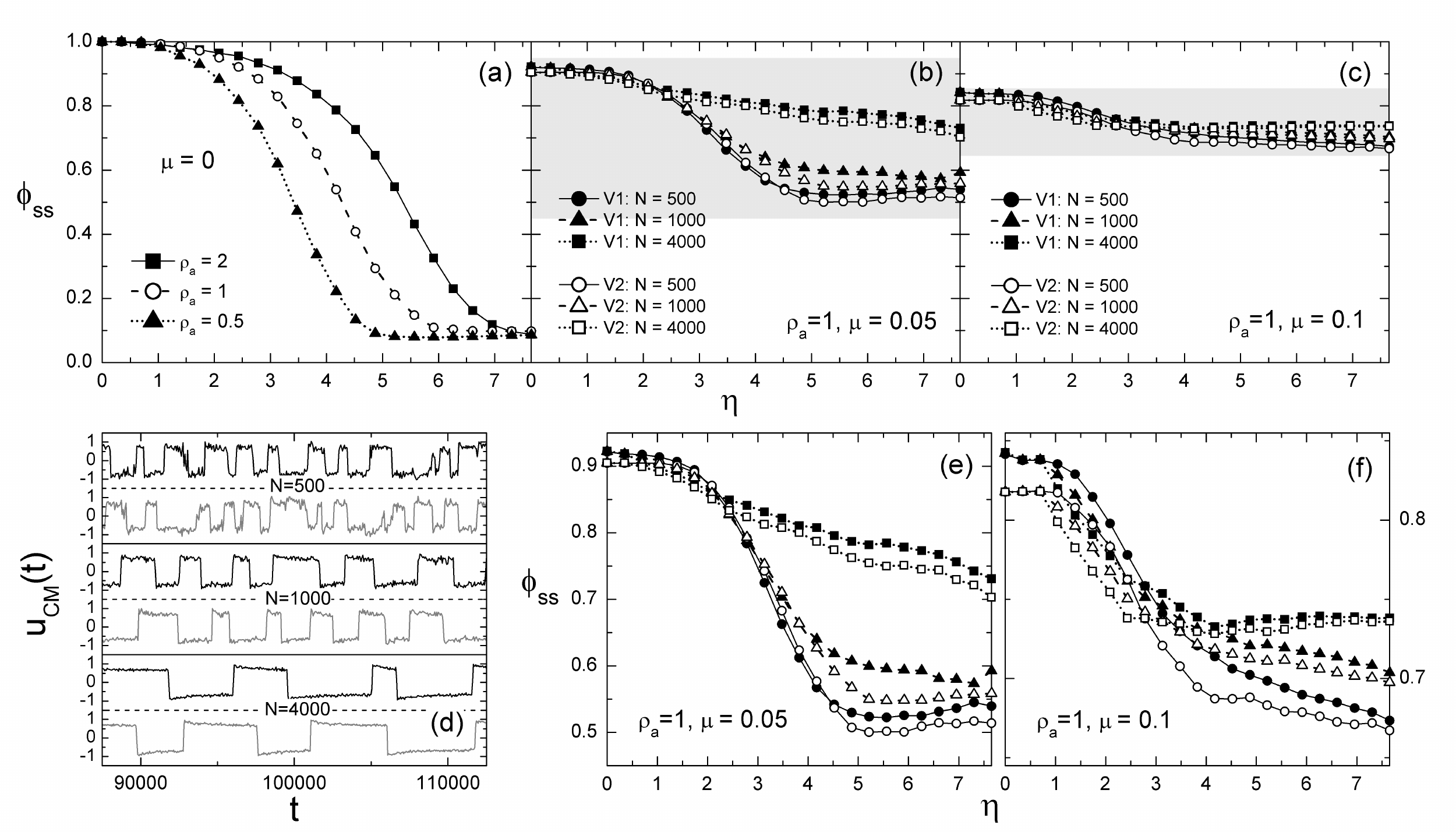} 
\caption{In (a), (b), (c), (e) and (f) the steady-state accumulated order parameter $\phi_{\rm ss}$, given in (\ref{eq:phiss}), is plotted as a function of the noise amplitude $\eta$. In each case, the system was let to evolve over 50000 time steps until it reached the steady state, and the order parameter was accumulated over 250000 time steps after the transient. In (a) the 1D Czir\'ok \etal model, given in (\ref{eq:1DVicsek1}) in its original form, is considered for different values of $\rho_{\rm a}$ and $N=1000$. In (b) and (c), and the details (e) and (f) corresponding to the grayed out portions of (b) and (c), respectively, Version 1 (V1) and Version 2 (V2) of our model, given in (\ref{eq:Version1}) and (\ref{eq:Version2}), are considered for $\rho_{\rm a}=1$, $\mu=0.05,0.1$ and different system sizes ($N=500, 1000, 4000$). In (d) the instantaneous velocity of the centre of mass $u_{\rm CM}(t)$, given in (\ref{eq:instmnvel}), is plotted for $\rho_{\rm a}=1$, $\mu=0.1$, $\eta=7$ and different system sizes in the steady state: the black solid lines correspond to Version 1 of our model while the grey solid lines to Version 2 (see the text for more details).}
\label{fig:ordpam}
\end{center}
\end{figure}

In order to quantify the time evolution of the system, we define the instantaneous order parameter as
\begin{equation}
\phi(t)=\frac{1}{N}\left|\sum_{i=1}^{N}u_i(t)\right|.
\label{eq:instordp}
\end{equation}
We also measured the instantaneous position and velocity of the centre of mass, given by
\begin{equation}
x_{\rm CM}(t)=\frac{1}{N}\sum_{i=1}^{N}x_i(t),
\label{eq:instxcm}
\end{equation}
and
\begin{equation}
u_{\rm CM}(t)=\frac{1}{N}\sum_{i=1}^{N}u_i(t),
\label{eq:instmnvel}
\end{equation}
respectively.

Before we proceed, there is one thing that needs to be clarified. Due to periodic boundary conditions, a 1D system has the topology of a ring. This implies there are two distances (running in opposite directions) from any particle to any specified point in the ring. Additionally, let us not forget that random initial conditions are taken both in the position and in the initial direction of the particles velocity, and that alignment interactions favour the formation of clusters as shown in figure \ref{fig:v1d-rho1}. For simplicity, the position of the centre of mass, as given in (\ref{eq:instxcm}), is determined with the particles position, $x_i(t)$, measured with respect to the origin from which the length $L$ of the system is measured as well. However, this does not affect the overall results obtained as we will see below.

To account quantitatively for the transition from the ordered to the disordered state, in figure \ref{fig:ordpam}(a) we plot the steady-state accumulated order parameter, defined as
\begin{equation}
\phi_{\rm ss}=\lim_{T\rightarrow\infty} \frac{1}{T} \int_0^T \phi(t) \, dt,
\label{eq:phiss} 
\end{equation}
as a function of the noise amplitude $\eta$ for $\rho_{\rm a}=0.5,1, 2$. This figure makes evident the dependence of the critical point $\eta_c$ on the average density of the system $\rho_{\rm a}$ (see more details in \cite{Czi99}).

\begin{figure}[t]
\begin{center}
\includegraphics[width=0.6\textwidth]{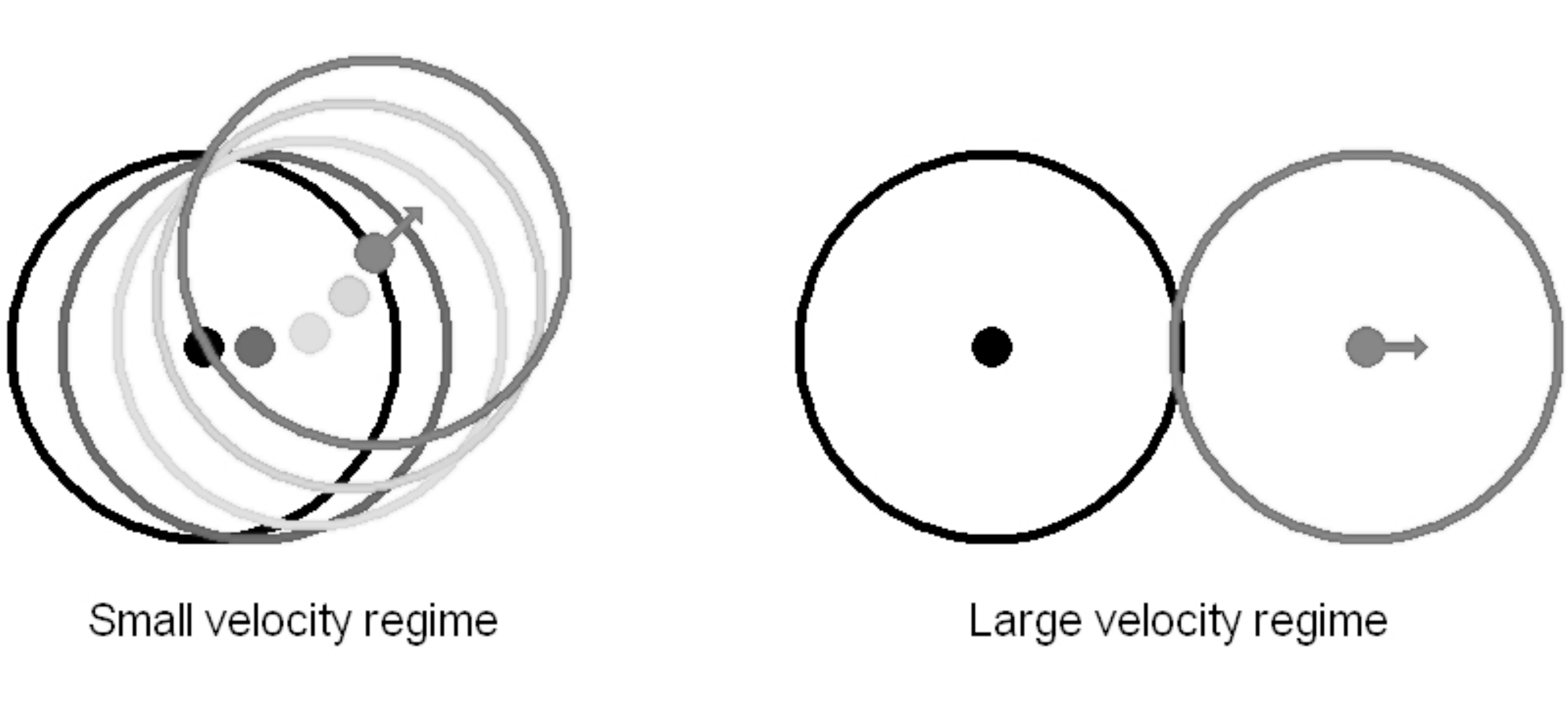} 
\caption{Schematic diagram of the \emph{small velocity regime} compared to the \emph{large velocity regime}.}
\label{fig:smallvr}
\end{center}
\end{figure}

Let us now provide the 1D Czir\'ok \etal model with a new interaction rule that considers \emph{centering} along with \emph{alignment}. However, there are some constraints that have to be taken into account. The dynamics in the Czir\'ok \etal model may be sensitive to the prescribed velocity $v_0$ of the particles. As reported in \cite{Nag07} for the original two-dimensional Vicsek \etal model \cite{Vic95} --- also known as the scalar noise model (SNM) --- an important quantity to be considered is the ratio of the interaction radius and the velocity of the particle, $R/v_0$, compared to the update time $\Delta t$ in the numerical simulations. The SNM model was proposed to study the motion of bird flocks and/or bacterial colonies. The motion of the particles in such systems is quasi-continuous, i.e., the reaction time of the birds is, usually,  significantly faster than the characteristic time that is needed to travel through their interaction radius $R$. This condition imposes the following constraint: $\Delta t \ll R/v_0$. In our case, once fixed the interaction radius $\Delta=1$ and the update time $\Delta t=1$, the above condition becomes $v_0\ll1$, the same as in the SNM model. This velocity domain is referred as the \emph{small velocity regime} (see figure \ref{fig:smallvr}). If the velocity of the particles is allowed to become larger ($v_0\ge0.3$, referred as the \emph{large velocity regime} in  \cite{Nag07}), the SNM model shows \emph{density waves} that appear due to an emergent influence of the periodic boundary conditions. Moreover, for very large particle velocites (e.g., $v_0=10$) the order-disorder transition exhibits a discontinuous order parameter characteristic of a first order phase transition (see \cite{Nag07}).

As far as we know, the properties of the Czir\'ok \etal model, as given in (\ref{eq:1DVicsek1}) and (\ref{eq:1DVicsek2}), have not yet been studied in detail outside the \emph{small velocity regime}. Nonetheless, it is known that for certain combination of the parameters, this model can develop directional switching (see, e.g., \cite{Bod10}).

Taking into account the previous statements, we will keep the system in the \emph{small velocity regime} ($v_0\ll1$), as we add \emph{centering} to the type of interactions experienced by the particles. In this way, on top of the majority rule that allows particles to match the velocity of their neighbours (interacting in the velocity space), we will introduce a simple attractive long-range rule (an interaction in the position space) that resembles that one used in the Mikhailov-Zanette model \cite{Mik99}. However, in contrast with the pairwise linear forces that attach the particles in the Mikhailov-Zanette model, in order to determine the new direction for a given particle, our \emph{centering-rule} for the Czir\'ok \etal model will only take into account the shortest distance from the position of that particle relative to the position of the centre of mass of the group $x_{\rm CM}$; this is done so because of the ring topology of the system. Additionally, to make the implementation of this centering-rule simple, the mean velocity of the particle's neighbours (obtained through function $G(\left<u(t)\right>_i)$) will then be directed towards the centre of mass whenever this rule is applied: we will call this function $H(\left<u(t)\right>_i;x_{\rm CM})$. In this way, we will be able to keep the velocity of the particles within the small velocity regime, so they will not be allowed to abandon their interaction regions in less than the characteristic time defined by the quantities $\Delta$, $v_0$, and $\Delta t$. This will also allow us to define two versions of the Czir\'ok \etal model, that include centering, in which these two rules are applied taking inspiration from two of the approaches to include stochastic errors as classified by Bode \etal \cite{Bod10}, listed in the Introduction of this work.

In Version 1 (V1) of our model we consider a probabilistic approach to apply the alignment and centering rules --- similar to approach ($ii$) according to Bode \etal (see above) --- but with synchronous updates. To weight one rule above the other we introduce the \emph{centering parameter} $\mu$ in the following way:
\begin{itemize}
\item With probability ($1-\mu$), a given bird in the flock will follow the original majority rule $G(\left<u(t)\right>_i)$, trying to match its velocity with the mean velocity of its neighbours (itself included).
\item With probability $\mu$, it will acquire the magnitude of the mean velocity of its neighbours, but will move towards the centre of mass of the flock instead, following rule $H(\left<u(t)\right>_i;x_{\rm CM})$ as described above.
\end{itemize}
We also keep the additive noise $\xi_i$ as in the original Czir\'ok \etal model. Thus, in this version of our model, the coordinate $x_i$ and dimensionless velocity $u_i$ of the particles are updated as
\begin{equation}
\eqalign{x_i(t+1) = x_i(t) + v_0 u_i(t), \cr
u_i(t+1) = I(\mu) + \xi_i,}
\label{eq:Version1}
\end{equation}
where $I$ is replaced by rule $G(\left<u(t)\right>_i)$ or $H(\left<u(t)\right>_i;x_{\rm CM})$ according to the value of $\mu$ in an probabilistic way.

In Version 2 (V2) of our model only the additive noise $\xi_i$ is kept as a source of stochasticity, applying the rules in a deterministic fashion like in the model of Buhl \etal \cite{Buh06}. In consequence, in this version of our model, the coordinate $x_i$ and dimensionless velocity $u_i$ of the particles are updated as
\begin{equation}
\eqalign{x_i(t+1) = x_i(t) + v_0 u_i(t), \cr
u_i(t+1) = (1-\mu) G(\left<u(t)\right>_i) + \mu H(\left<u(t)\right>_i;x_{\rm CM}) + \xi_i.}
\label{eq:Version2}
\end{equation}
In this way, in both versions of our model one can go from the original Czir\'ok \etal model (for $\mu=0$) to one that resembles the interaction type present in the Mikhailov-Zanette (for $\mu>0$). This allows to continuously go from a purely alignment type of model to a purely centering one (for $\mu=1$). However, let us not forget that the scaling in the Czir\'ok \etal model depends on the average density of the particles $\rho_{\rm a}$ and the size of the system $L$, characteristic not shared by the Mikhailov-Zanette model where none of its properties depend on the density of particles.

\begin{figure}[tbp]
\begin{center}
\includegraphics[width=\textwidth]{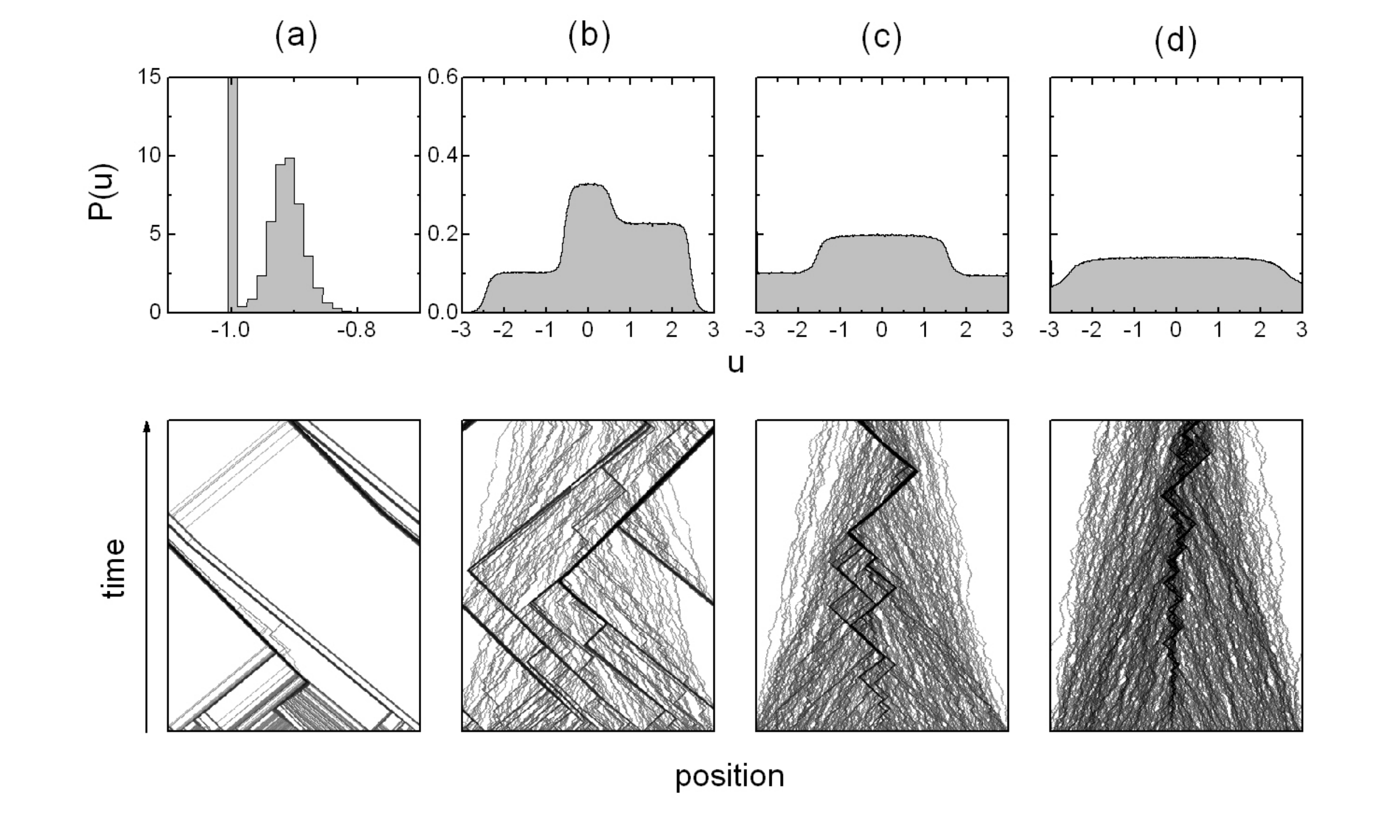} 
\caption{The dynamics of Version 1 of our model, given in (\ref{eq:Version1}), is presented in the  lower plots for $L=1000$, $N=1000$, $\rho_{\rm a}=1$ and $\mu=0.05$, for increasing values of $\eta$ (from left to right $\eta=0,3,5,7$, respectively). The darker grey level represents higher particle density and the systems were let to evolve 15000 time steps. The upper plots present the corresponding normalized distribution $P(u)$ of the velocities $u_i$ of the individual particles in the steady state. Please notice how, in the absence of noise ($\eta=0$), $P(u)$ shows two peaks centered (from left to right) around the velocity of particles not yet absorbed by the densest cluster and around the velocity of the densest cluster, respectively (see the text for more details).}
\label{fig:v1d-mu05rho1}
\end{center}
\end{figure}

\section{Results and discussion}
We will now consider the effects of centering and alignment in the Czir\'ok \etal model by studying cases with $\mu>0$, first for Version 1 of our model in subsection~\ref{ssec:condstates}. Later we show in subsection~\ref{ssec:noise} that, aside from some minor differences, Version~1 and Version~2 of our model are qualitatively equivalent regarding the collective behaviour of the system. We also contrast our results with those for some other models. Finally, in subsection~\ref{ssec:scaling} we present the scaling properties for both versions of our model.

\subsection{Condensed states and directional switching
\label{ssec:condstates}}

In the lower plots of \hbox{figure \ref{fig:v1d-mu05rho1},} we show the time evolution (i.e., the evolution of $\rho(x,t)$) of Version 1 of our model given in (\ref{eq:Version1}), for $L=1000$, $N=1000$ and $\mu=0.05$, and increasing values of the noise amplitude $\eta(\in[0,7])$ from right to left, respectively. The symmetry breaking can still be appreciated as shown in the lower plots of figures \ref{fig:v1d-mu05rho1}(a) and \ref{fig:v1d-mu05rho1}(b). By increasing the noise amplitude, $\eta$, the system is driven from a symmetry-broken ordered phase to an \emph{alternating flock} phase, shown in the lower plots of figures \ref{fig:v1d-mu05rho1}(c) and \ref{fig:v1d-mu05rho1}(d). It is very clear the effect of the centering rule, eventually gathering all of the particles in one big cluster. Moreover, one can appreciate how the cluster with highest density (darkest in the figures), and thus always close to the centre of mass,  develops a slower mean velocity with respect to the rest of the clusters: a close inspection of the slope of this cluster accounts for it, with its slope being higher compared to the slope of other clusters. This can be better appreciated by looking at the distribution $P(u)$ of the velocities of the individual particles, shown in the upper plots of figure \ref{fig:v1d-mu05rho1}. In particular, notice the bimodal distribution of figure \ref{fig:v1d-mu05rho1}(a), where the narrow peak centered at $-1$ corresponds to those particles not yet absorbed by the densest cluster, while the wider one, centered a little to the left of $-0.9$, corresponds to the latter. As a result, the order parameter in the steady state, $\phi_{\rm ss}$, never reaches a value equal to $1$, even in the absence of noise ($\eta=0$). On the other hand, as the noise intensity increases, $\phi_{\rm ss}$ never vanishes as the system develops a mean velocity with a positive magnitude in the alternating flock phase and contrary to what happens in the absence of centering. See, e.g., figure \ref{fig:ordpam}(b) for $\rho_{\rm a}=1$ and $\mu=0.05$ where different system sizes are considered (lines with solid symbols).

\begin{figure}[tbp]
\begin{center}
\includegraphics[width=\textwidth]{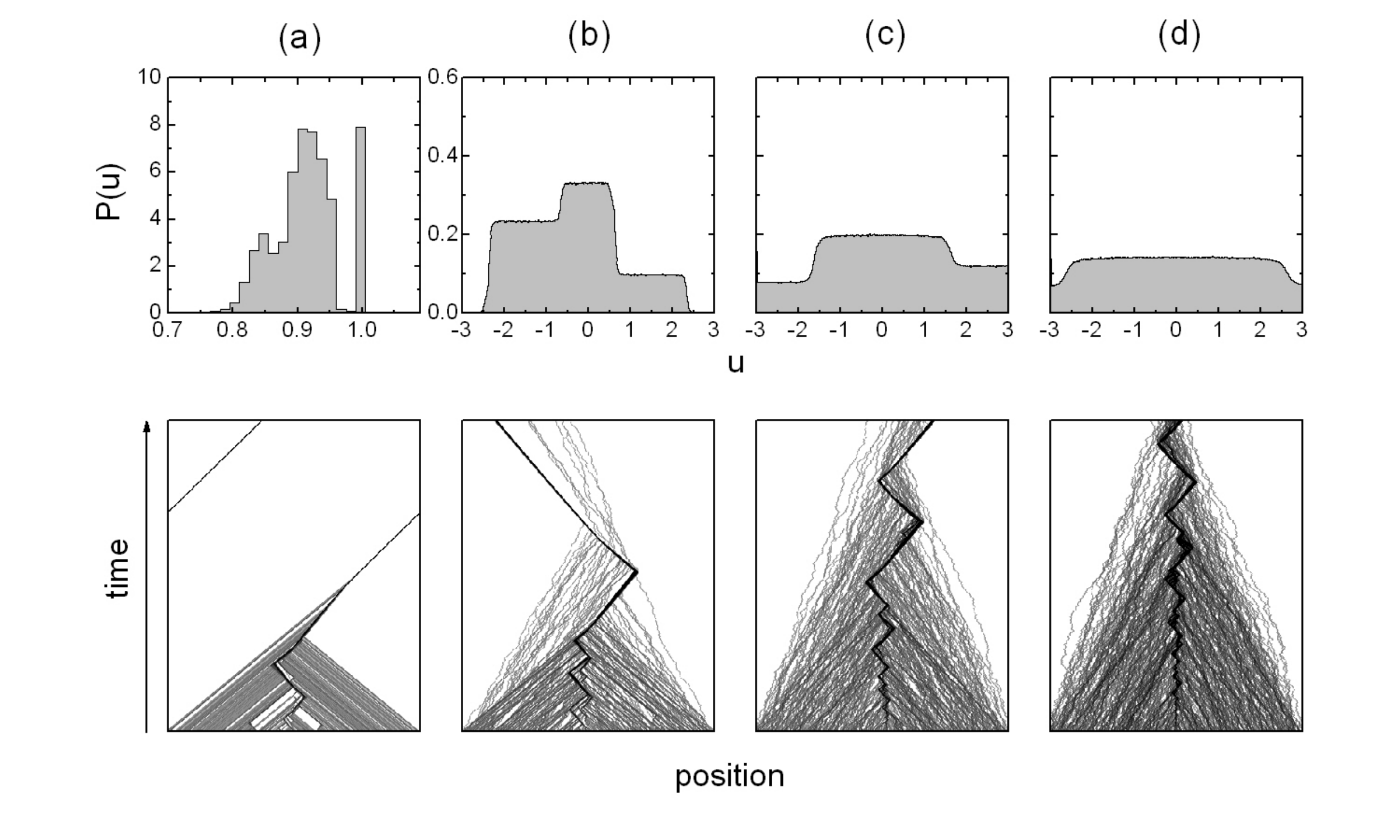}
\caption{The dynamics of Version 1 of our model, given in (\ref{eq:Version1}), is presented in the  lower plots for $L=1000$, $N=1000$, $\rho_{\rm a}=1$ and $\mu=0.1$, for increasing values of $\eta$ (from left to right $\eta=0,3,5,7$, respectively). The darker grey level represents higher particle density and the systems were let to evolve 15000 time steps. The upper plots present the corresponding normalized distribution $P(u)$ of the velocities $u_i$ of the individual particles in the steady state. As $\eta$ increases, the frequency of changes in the direction of motion of the densest cluster increases as well.}
\label{fig:v1d-mu1rho1}
\end{center}
\end{figure}

These effects are only enhanced as $\mu$ increases. For $\mu=0.1$ and $\eta=0$, the order parameter, $\phi_{\rm ss}$, reaches a smaller value in comparison with the case $\mu=0.05$. In contrast, for strong values of the noise intensity $\eta$, the alternating flock phase develops an order parameter with higher values than for $\mu=0.05$. See the lines with solid symbols in figure \ref{fig:ordpam}(c) ($\rho_{\rm a}$ was kept constant in both cases).

To better illustrate this, the lower plots in figure \ref{fig:v1d-mu1rho1} show the dynamics of Version 1 of our model with $\mu=0.1$. As expected, as $\mu$ increases, both condensed phases are reached faster. Yet again, the densest cluster develops a slower mean velocity than the ``free'' particles that have not yet been absorbed by it, shown by the double-peaked distribution in the upper figure \ref{fig:v1d-mu1rho1}(a), with a wide peak centered a little less than $0.9$ and a narrow one centered at $1$, respectively. As noise increases, for strong values of the noise intensity $\eta$, the alternating flock phase becomes more evident as shown in the lower plots of figures \ref{fig:v1d-mu1rho1}(c) and \ref{fig:v1d-mu1rho1}(d). As apparent from the figures, the frequency of the changes in direction of the densest-cluster mean velocity increases with $\eta$, while the distribution $P(u)$ broadens out (see the upper plots in figure \ref{fig:v1d-mu1rho1}).

\begin{figure}[tbp]
\begin{center}
\includegraphics[width=\textwidth]{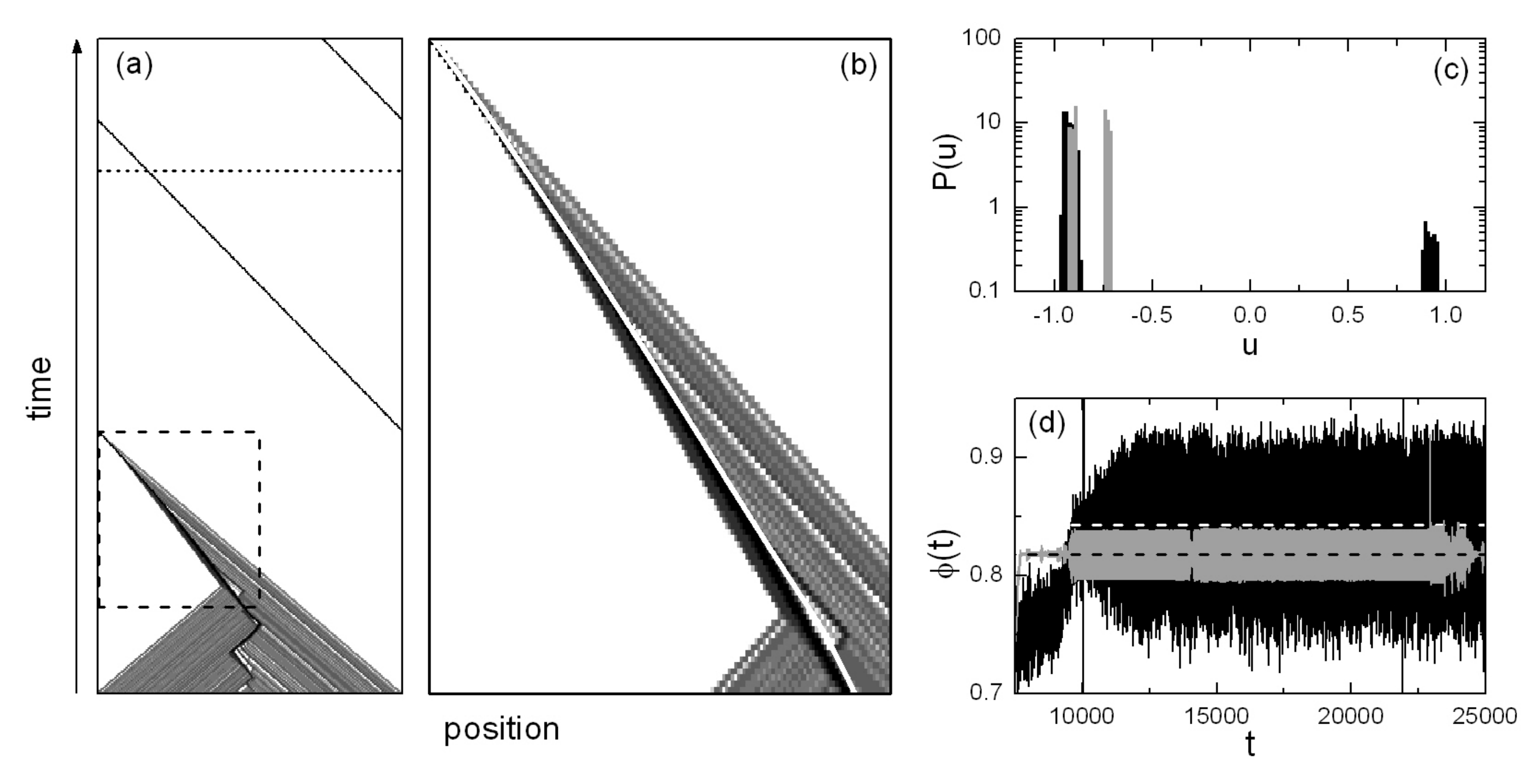}
\caption{(a) The dynamics of Version 1 of our model, given in (\ref{eq:Version1}), is presented for $L=1000$, $N=1000$, $\rho_{\rm a}=1$, $\mu=0.1$ and $\eta=0$. The darker grey level represents higher particle density. The system was let to evolve 25000 time steps, while the steady state measurements were taken above the horizontal dotted line corresponding to $t=20000$. \hbox{(b) Detail} of the dynamics, marked with a dashed square in (a). The trajectory of the centre of mass, $x_{\rm CM}$, is depicted with a white solid line. \hbox{(c) Semi-log} plot of the normalized distribution $P(u)$ of the velocities of the individual particles, in the steady state, for Version 1 (solid black) and Version 2 (solid grey) of our model with the same parameters as in (a); Version 2 of our model is given in (\ref{eq:Version2}). The peaks of $P(u)$ for each version of our model are centered differently (symmetric vs.\ non-symmetric) due to the way alignment and centering interactions are implemented (see the text for more details). \hbox{(d) Instantaneous} order parameter $\phi(t)$, given in (\ref{eq:instordp}), for Version 1 (solid black curve) and Version 2 (solid grey curve) of our model. The horizontal white and black dashed lines correspond to the order parameter $\phi_{\rm ss}$, for Version 1 and Version 2 of our model, respectively, accumulated in the steady state with the same parameters as in (a). Due to the extra source of stochasticity, in the probabilistic way alignment and centering interactions are implemented in Version 1 of our model, the fluctuations of the instantaneous order parameter $\phi(t)$ are stronger than in the deterministic Version 2.}
\label{fig:m1e0ss}
\end{center}
\end{figure}

In order to explain this, let us consider the combined effects of the alignment and centering interactions on the densest cluster, first for the case without noise ($\eta=0$). Because of the presence of centering, particles in this cluster consistently move towards the centre of mass even against the direction of motion of the majority. This results in a slowdown of the mean velocity of the group.\footnote{Please remember that the alignment interaction determines the instantaneous velocity of a given particle from the local mean velocity of its neighbours through function $G(u)$ as $G(\left<u(t)\right>_i)$.} See, e.g., the steady state of the case $\mu=0.1$, where an ordered phase with broken symmetry develops, moving to the left as shown in \hbox{figure \ref{fig:m1e0ss}(a)}. For particles at the front of the densest cluster, the centre of mass lies behind as illustrated with the white solid line in the detail given in figure \ref{fig:m1e0ss}(b). In consequence, particles at the front turn back to the right from time to time, opposing the direction of the majority. However, the alignment interaction keeps particles moving to the left most of the time, switching the direction of the opposing particles once again. This means that any given particle moves mainly in the direction of motion of the group, but it also moves in the opposite one as corroborated by looking at $P(u)$ in figure \ref{fig:m1e0ss}(c), depicted with the solid black curve for this case. There, the higher peak is centered around the emergent mean velocity of the group, while the shorter one (in a symmetric position) appears due to the opposing particles. 

\begin{figure}[tbp]
\begin{center}
\includegraphics[width=\textwidth]{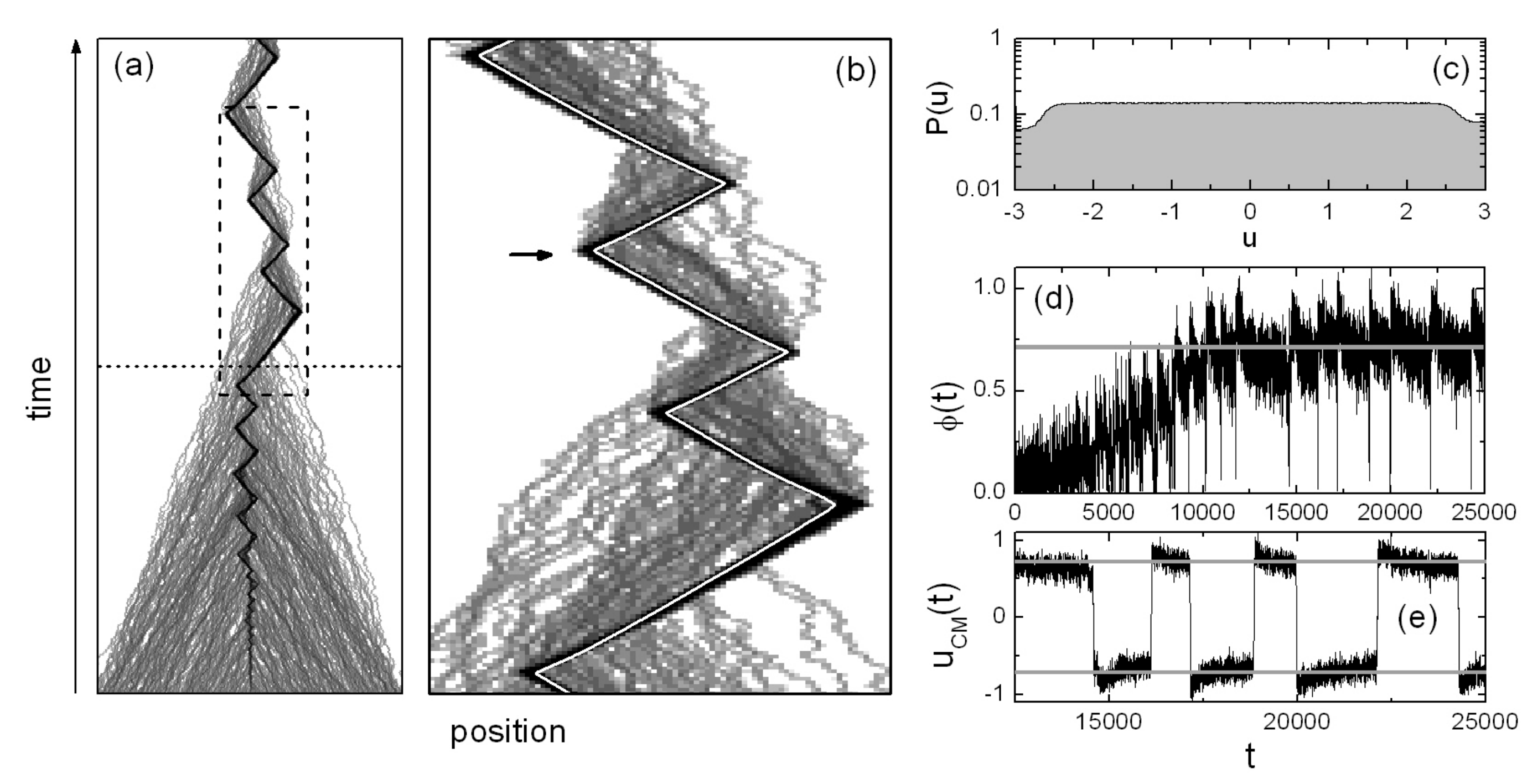}
\caption{(a) The dynamics of Version 1 of our model, given in (\ref{eq:Version1}), is presented for $L=1000$, $N=1000$, $\rho_{\rm a}=1$, $\mu=0.1$ and $\eta=7$. The darker grey level represents higher particle density. The system was let to evolve 25000 time steps, while the steady state measurements were taken above the horizontal dotted line corresponding to $t=15000$. \hbox{(b) Detail} of the dynamics, marked with a dashed square in (a). The trajectory of the centre of mass, $x_{\rm CM}$, is depicted with a white solid line. Notice how, on one side, the densest cluster effectively oscillates around the center of mass, while a tail of defecting particles counterbalances its mass on the other. \hbox{(c) Semi-log} plot of the normalized distribution $P(u)$ of the velocities of the individual particles in the steady state. \hbox{(d) Instantaneous} order parameter $\phi(t)$, given in (\ref{eq:instordp}), for the whole time evolution of the system. \hbox{(e) Instantaneous} velocity of the centre of mass $u_{\rm CM}(t)$, given in (\ref{eq:instmnvel}), for the steady state of the system. In (d) and (e), the horizontal grey lines corresponds to the accumulated mean value $\phi_{\rm ss}$, and $\pm\phi_{\rm ss}$, respectively.}
\label{fig:m1e7ss}
\end{center}
\end{figure}

In this way, once a cluster is formed around the centre of mass, this will grow denser over the rest of the systems' time evolution, gathering more particles together and, at the same, hindering their ability to fully develop the preferred velocity as dictated by function $G(u)$, given in (\ref{eq:1DVicsek2}). In some sense, inside the densest cluster, the combination of centering and alignment works as a source of ``noise'' itself (also established in \cite{Bod10}), inducing fluctuations in the steady-state mean velocity of the group even for $\eta=0$. See, e.g., \hbox{figure \ref{fig:m1e0ss}(d)} where the instantaneous order parameter $\phi(t)$ is plotted as a function of time (solid black curve). The horizontal white dashed line corresponds to the value of $\phi_{\rm ss}$ measured in the steady state, above the horizontal dashed line shown in figure \ref{fig:m1e0ss}(a). Regarding the ``free'' particles not yet absorbed by the densest cluster, in the early stages of evolution, the effect of centering on their speed is negligible compared to the alignment interaction, due to the much lower density in these regions. In the end, the magnitude of the emergent velocity of the densest cluster, when the noise is weak or absent, will depend on the balance of the centering and alignment interactions.

\begin{figure}[t]
\begin{center}
\includegraphics[width=\textwidth]{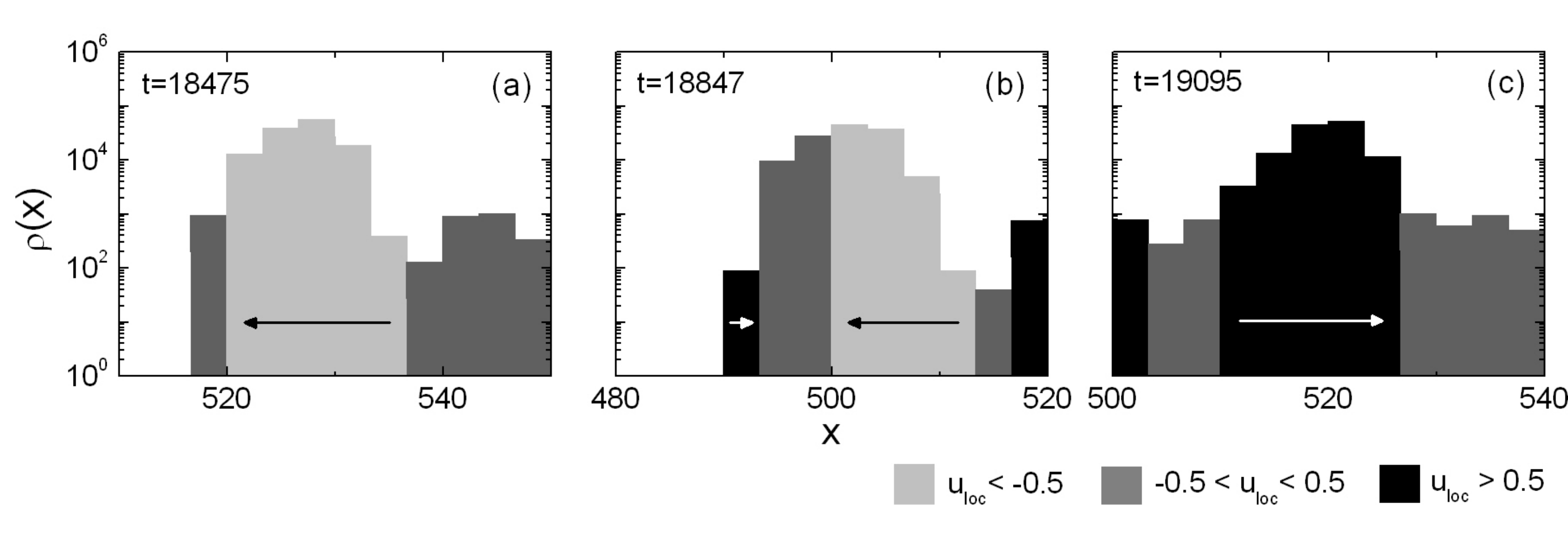}
\caption{Semi-log plots of the density of particles $\rho$, as a function of the position $x$, taken over $62$ consecutive times steps. In (a), (b) and (c), three cases are plotted \emph{before}, \emph{at}, and \emph{after} the turn marked with a small arrow in figure \ref{fig:m1e7ss}(b), around $t=18475, 18847, 19095$, respectively. The levels of grey correspond to the local mean velocity, $u_{\rm loc}$, as shown in the scale at the bottom-right, while the horizontal black and white arrows point in its direction. Notice the development of domains with different local mean velocity.}
\label{fig:m1e7rho2}
\end{center}
\end{figure}

Let us now consider the case when the noise is strong. Here again, the combined effects of alignment and centering are stronger on the densest cluster. In figure \ref{fig:m1e7ss}(a), the evolution of $\rho(x,t)$ is shown for a system with $\mu=0.1$. One can appreciate that, as soon as a large and dense cluster is formed, it starts to change its direction of motion with some periodicity. In consequence, the system develops an alternating flock phase that resembles that in the models of Evans \etal \cite{Olo99,Ray06}. To quantify this, it is better to look at the instantaneous velocity of the centre of mass, $u_{\rm CM}$, given in (\ref{eq:instmnvel}). As shown in figure \ref{fig:m1e7ss}(e), in the steady state, the velocity of the centre of mass develops a constant mean magnitude (see the horizontal grey lines) but with an alternating direction. In contrast with the zero-noise case, due to the presence of noise ($\eta>0$), particles are able to drift away from the centre of mass in a ``diffusive'' fashion. This can be observed when looking at the steady state of the system, above the horizontal dotted line in figure \ref{fig:m1e7ss}(a). As more particles leave the densest cluster, this becomes dilute and wider as the centre of mass is ``pulled'' to its back as shown by the solid white line in the detail plotted in figure \ref{fig:m1e7ss}(b). The turn marked with the small arrow in this figure is analyzed in more detail in figure \ref{fig:m1e7rho2}. There, the density of particles (accumulated over $62$ consecutive time steps) is plotted as a function of the position around three different configurations: \emph{before}, \emph{at}, and \emph{after} the turn. It is clear the development of domains with different local mean velocity, $u_{\rm loc}$, corresponding to the different levels of grey. As shown in figures \ref{fig:m1e7rho2}(a) and \ref{fig:m1e7rho2}(c), before and after the turn, when the centre of mass is closer to the densest cluster (see figure \ref{fig:m1e7ss}(b)), the latter is surrounded by domains with small $u_{\rm loc}$. However, at the turn, the densest cluster is at its farthest position relative to the position of the centre of mass (see figure \ref{fig:m1e7ss}(b)). A fluctuation develops at the front of the flock, followed by a small-$u_{\rm loc}$ domain as shown in figure \ref{fig:m1e7rho2}(b). This fluctuation moves against the direction of the flock, gaining density and momentum, finally inverting the direction the flock as it passes through. This mechanism effectively renders the particles in the system oscillating around the centre of mass, with the densest cluster on one side, and the defecting particles counterbalancing its mass on the other as can be appreciated in figure \ref{fig:m1e7ss}(b). In this way, the alternating flock phase in our model resembles the oscillatory phase in the Mikhailov and Zanette model \cite{Mik99}, that also considers long-range centering but not alignment.

Two conclusions can be drawn from this analysis. The first relates the frequency of the turns with the intensity of the noise. As explained before, the presence of noise allows particles to drift away from the densest cluster that always remains close to the centre of mass. The more intense the noise is, the faster particles are able to move away. Consequently, the frequency of the turns should increase with the noise amplitude as figures \ref{fig:v1d-mu1rho1}(c) and \ref{fig:v1d-mu1rho1}(d) can confirm, a trend that is consistent with previous results; see, e.g., \cite{Yat09, Olo99, Bod10}. The second regards the development of a mean velocity different from zero by the densest cluster when strong noise is present. In contrast with the case without centering ($\mu=0$), where strong noise makes the mean velocity of the flock vanish in a continuous phase transition, when centering is introduced ($\mu>0$) the accumulated order parameter $\phi_{\rm ss}$ is far from zero -- it even increases with $\mu$!

For strong noise, centering in combination with the alignment interaction provokes an ordering effect on the densest cluster. This can be understood considering that, due to centering, particles try to move towards the centre of mass of the group eventually gathering together. With the aid of the alignment interaction, correlations in the velocities $u_i$ of the particles develop that cannot be destroyed by simple additive noise. This is due to the long range character of the centering interaction and the high density of this cluster. In contrast, the alignment interaction being short-ranged in nature, is more sensitive to the noise when taken alone. This may also explain the increase of $\phi_{\rm ss}$ with $\mu$, being centering the dominant interaction for strong noise.

\begin{figure}[t]
\begin{center}
\includegraphics[width=\textwidth]{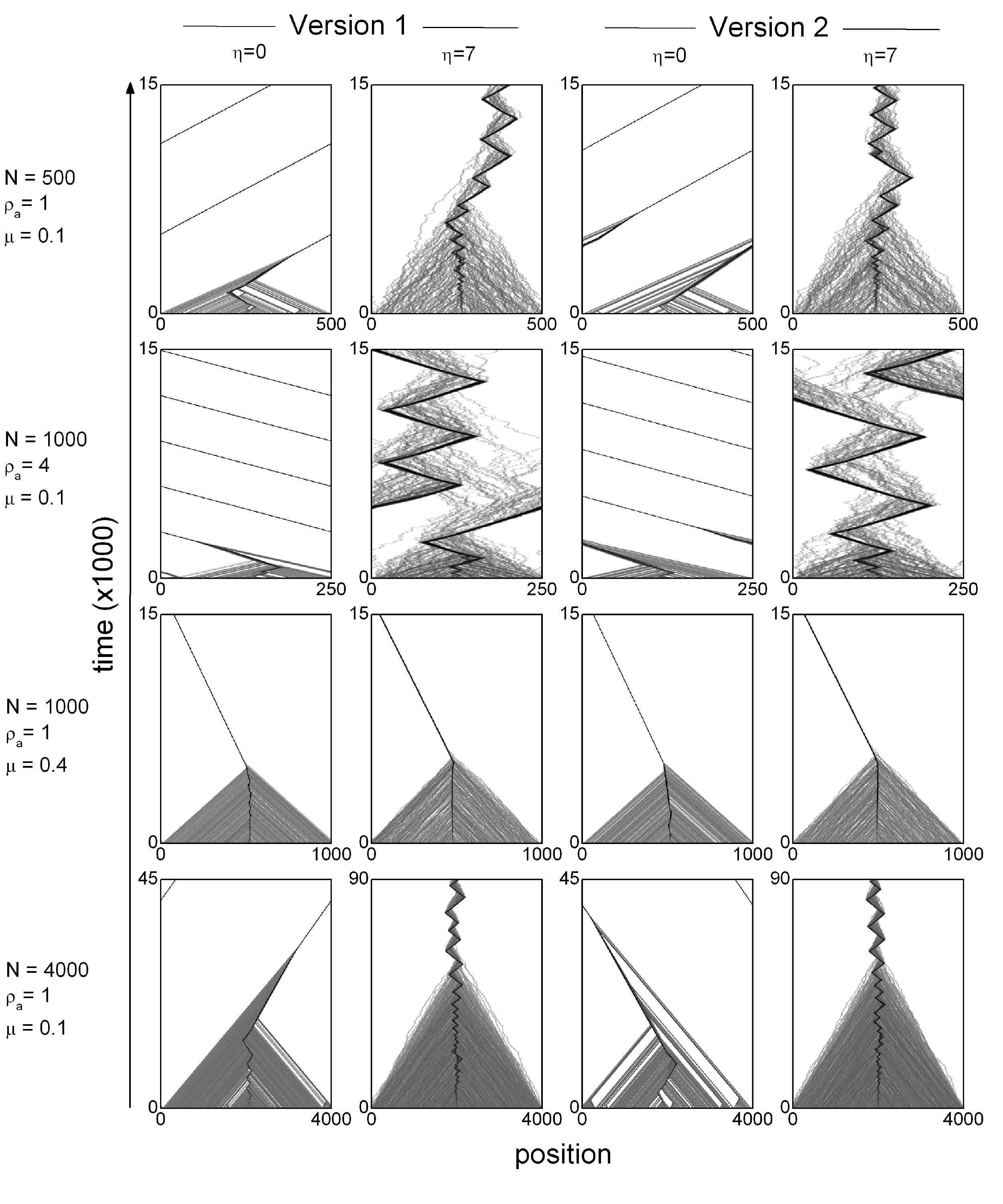}
\caption{Dynamics of Version 1 and Version 2 of our model for different combinations of the parameters $N$, $\rho_{\rm a}$ and $\mu$ in two limit cases: $\eta=0$ and $\eta=7$. The darker grey level represents higher particle density. The case on the third row is of particular interest since the alternating flock phase is fully suppressed even for strong noise (see the text for more details).}
\label{fig:sixteen}
\end{center}
\end{figure}

\subsection{Different sources of stochasticity and free flocks
\label{ssec:noise}}

Regarding the collective behaviour of the system, Version 2 of our model shows the same overall features as Version 1. For that matter, the dynamics of four systems with different combinations of the parameters $N$, $\rho_{\rm a}$ and $\mu$, is presented in figure \ref{fig:sixteen} for both versions, side by side, in two limit cases: $\eta=0$ and $\eta=7$. These examples illustrate the qualitative equivalence (and, in great measure, quantitative too) between the two versions of our model --- as apparent from the figures --- including the alternating flock phase. The case $N=1000$, $\rho_{\rm a}=1$ and $\mu=0.4$ in the third row, is of particular interest since this phase is absent even for strong noise ($\eta=7$). We will discuss this case later.

\begin{figure}[tbp]
\begin{center}
\includegraphics[width=\textwidth]{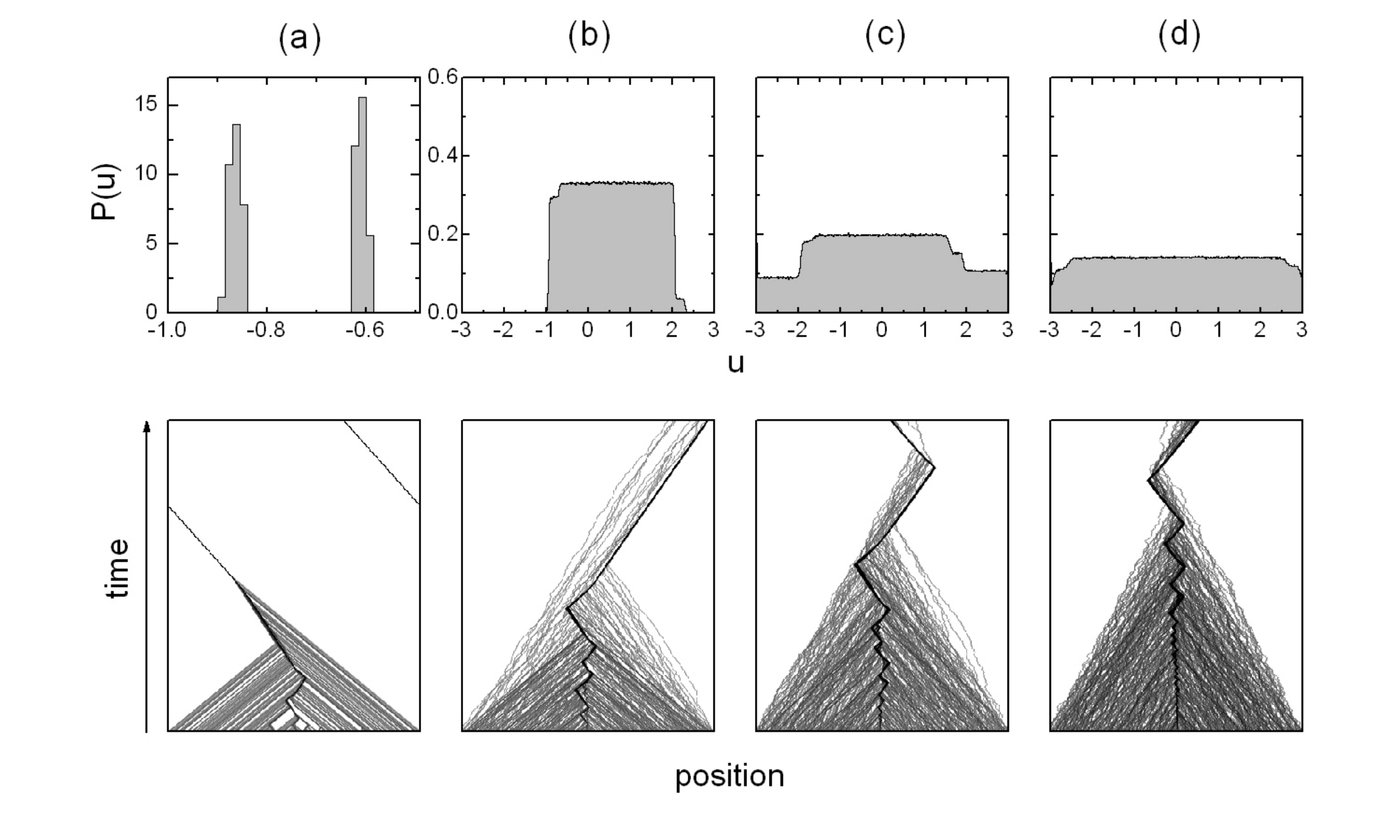}
\caption{The dynamics of Version 2 of our model, given in (\ref{eq:Version2}), is presented in the  lower plots for $L=1000$, $N=1000$, $\rho_{\rm a}=1$ and $\mu=0.15$, for increasing values of $\eta$ (from left to right $\eta=0,3,5,7$, respectively). The darker grey level represents higher particle density and the systems were let to evolve 15000 time steps. The upper plots present the corresponding normalized distribution $P(u)$ of the velocities $u_i$ of the individual particles in the steady state. Aside from the form of the distribution of $P(u)$, this version of our model shows the same overall features as Version 1.}
\label{fig:v1d-mu15rho1_v2}
\end{center}
\end{figure}

More details of the dynamics of Version 2 are presented in figure \ref{fig:v1d-mu15rho1_v2} for $\mu=0.15$, with the same values of $\rho_{\rm a}$, $N$, $L$ and $\eta$ as in the two cases presented for Version 1 in figures \ref{fig:v1d-mu05rho1} and \ref{fig:v1d-mu1rho1}. Again, the system shows collective behavior equivalent to that of Version 1. Moreover, the mechanism that produces the alternating flock phase in Version 2, is exactly the same as the one described in subsection \ref{ssec:condstates}. Nonetheless, there are some differences that need to be pointed out. The first regards the distribution $P(u)$ of the velocities of the individual particles for the case $\eta=0$, shown in the upper plot of figure \ref{fig:v1d-mu15rho1_v2}(a), as this exhibits two peaks centered around $-0.9$ and $-0.6$.\footnote{A third peak should appear around $-1$, corresponding to particles not-yet-absorbed by the densest cluster. This peak is not present here since there were no such particles when $P(u)$ was measured.} As in Version 1, in Version 2 of our model, particles moving in the direction of the densest cluster move with speeds close to $|1|$, in contrast, particles moving in the opposite direction move with speeds around $|1| - 2\mu$. These values correspond to the centers of the two peaks of $P(u)$ shown in figure \ref{fig:v1d-mu15rho1_v2}(a). This comes as a result of the deterministic way the alignment and centering rules are applied in Version 2; see (\ref{eq:Version2}). Also, please notice that these peaks are narrower and comparable in height, in contrast to the corresponding ones for Version 1 --- wider, symmetric in position, but highly asymmetric in height ---  as illustrated in figure \ref{fig:m1e0ss}(c) for both versions with the solid grey and black filled curves, respectively ($\mu=0.1$ in this case).

This comes about the fact that, in both versions of our model, the combination of alignment and centering is a source of noise in itself --- all the more so, considering the additive noise is absent for $\eta=0$. However, Version 1 of our model has an extra source of stochasticity in the probabilistic way these interactions are applied; see (\ref{eq:Version1}). As a result, the fluctuations of the instantaneous order parameter $\phi(t)$ are stronger in Version 1 than in Version 2, as shown in figure \ref{fig:m1e0ss}(d) for $\mu=0.1$ with the solid black and grey curves, respectively. Before we go on with the discussion, let us provide a context regarding the development of directional switching and its relation with different sources of stochasticity in 1D SPP models.

In \cite{Bod10}, Bode \etal state that the addition of simple noise terms in the particles direction of motion is not necessarily sufficient to describe and explain collective motion in animal groups. This remark is made in regard of the directional switching shown by the marching locusts in the experiments by Yates \etal reported in \cite{Yat09}. In the latter, a variation of the Czir\'ok \etal model is proposed to explain this behaviour, where the local alignment interaction competes with each particle's own velocity in addition to a non-trivial additive noise. In response, Bode \etal proposed a fully stochastic 1D SPP model that does not consider any kind of additive noise, but an intrinsic source of stochasticity coming from an asynchronous updating with random sampling of neighbours, along with local concentric centering and alignment interaction zones. From both versions of our model, one can appreciate that the stochasticity coming from the combination of alignment and long-range centering interactions (in a synchronous updating framework) is not enough for the system to develop directional switching (the alternating flock phase); see, e.g., all the cases with $\eta=0$ in figures  \ref{fig:v1d-mu05rho1}, \ref{fig:v1d-mu1rho1}, \ref{fig:m1e0ss}, \ref{fig:sixteen} and \ref{fig:v1d-mu15rho1_v2}. Even more, the extra stochasticity coming form the probabilistic application of these interactions in Version 1 is still not enough. Nonetheless, both versions of our model have built-in the capacity to develop directional switching, that gets unleashed as soon as an extra source of noise is added to the system. In our case, the simple additive noise term $\xi_i$ in (\ref{eq:Version1}) and (\ref{eq:Version2}) does the trick, which leads us to the following question: Why does the noisier Version 1 behave qualitatively in the same way as Version 2 regarding the collective behaviour of the system, including the directional switching, and the absence of it when the additive noise is not present? 

\begin{figure}[t]
\begin{center}
\includegraphics[width=\textwidth]{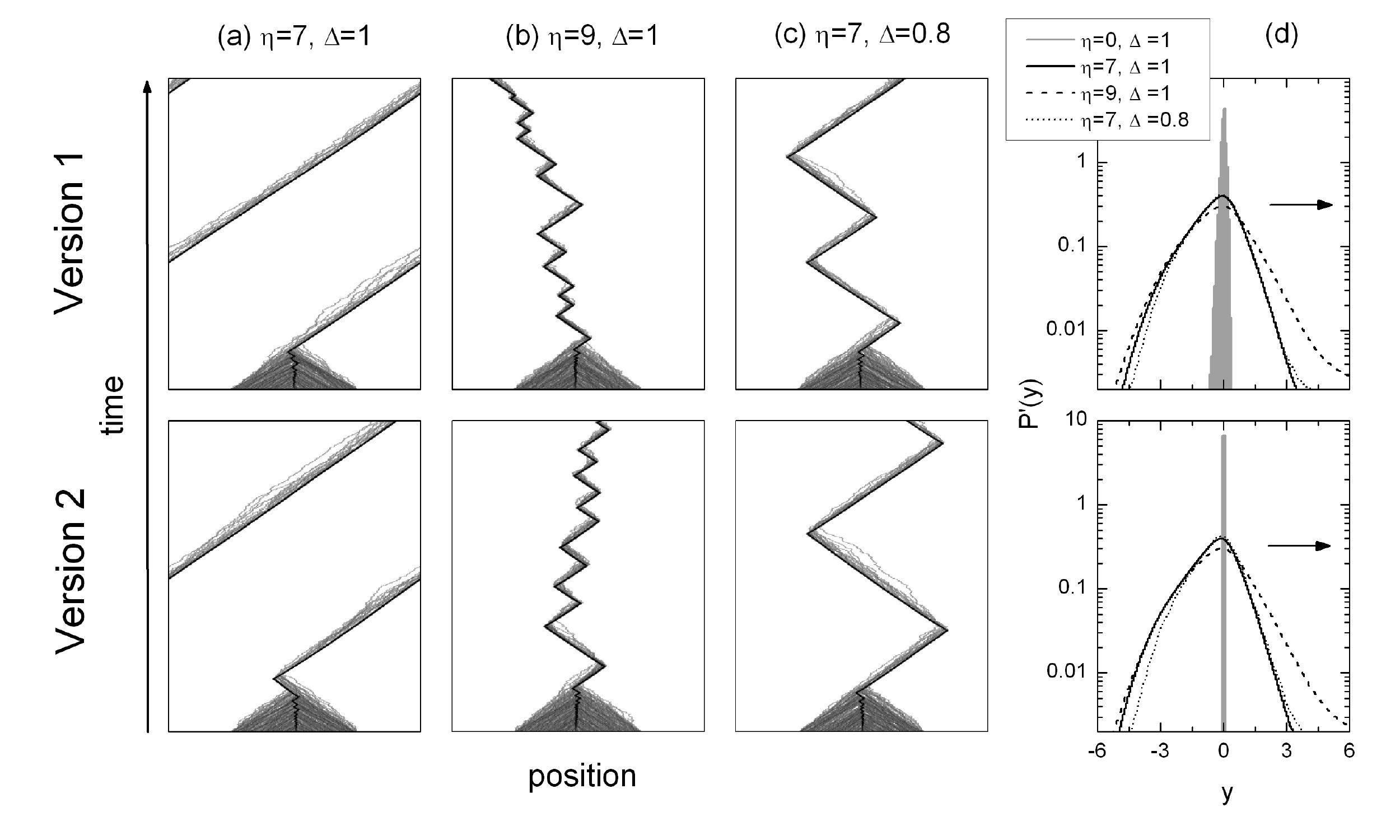}
\caption{In (a), (b) and (c) the dynamics of Version 1 and Version 2 of our model, for $N=1000$, $v_0=0.1$ and $\mu=0.2$, is presented for different values of the noise intensity $\eta$ and the alignment interaction radius $\Delta$. The darker grey level represents higher particle density. Periodic boundary conditions have been relaxed, as the particles are allowed to freely move in all of position space. They were only applied to the snapshots we show, in order to limit them to a finite region of space: in this case, $L=1000$. The systems were allowed to evolve 30000 time steps. In (a) $\eta=7$ and $\Delta=1$. In (b) $\eta=9$ and $\Delta=1$. In (c) $\eta=7$ and $\Delta=0.8$. (d) Semi-log plots of the normalized distribution $P'(y)$ of the individual particles distance to the centre of mass, measured in the steady state (starting after 10000 time steps of evolution) for the different cases (see the text for more details).}
\label{fig:free_m2}
\end{center}
\end{figure}

To answer this question, with an eye on the arguments provided above, we propose the study of a \emph{free} flock. For this, we will relax the periodic boundary conditions in both versions of our model, allowing the system to freely move in all of position space. We can do so, since the cohesion of our system is guaranteed by the long-range character of our centering interaction that contrasts with the local interactions considered in the previously mentioned models. This also leads us back to the case of the third row in figure \ref{fig:sixteen} where, even for the strong noise case ($\eta=7$ in the figure), the alternating flock phase is absent. In clarifying the origin of directional switching in our model, we hope to shed some light on the origin of this kind of state in the models discussed above, but before, there are some extra features of our model that must be mentioned:
\begin{itemize}
\item First, none of the properties of our model (in both of its versions) depend on the average particle density $\rho_{\rm a}$, but only on the number of particles $N$ as will be made clear in the next subsection. This is a consequence of the long-range character of the centering interaction that we have chosen, and resembles the properties of some other models with this kind of interactions (see, e.g., \cite{Mik99, Dos09}) --- in particular, the Mikhailov and Zanette model that inspired ours. \item Second, for values of $\mu>0.2$, both versions of our model are unable to develop an alternating flock phase in the steady state, at least for systems with $N=1000$ as shown in figure \ref{fig:free_m2}(a), where $\eta=7$ and $\Delta=1$.
\end{itemize}
This leaves us with the fact that, even with all of the sources of stochasticity active in Version 1 and Version 2 or our model as explained above, a strong-enough centering interaction can suppress the alternating flock phase. In this way, given a fixed number of particles $N$, we will consider two quantities to vary in order to recover it: the noise intensity $\eta$ and the interaction radius $\Delta$. Regarding the latter, let us not forget that the condition $v_0 \ll 1$ for the small velocity regime must be fulfilled ($\Delta t=1$ and $v_0=0.1$ are kept as before).

In figure \ref{fig:free_m2}(b), we have increased the amplitude of the noise from $\eta=7$ to 9 compared to the case of figure \ref{fig:free_m2}(a). As shown in the figure, the alternating flock phase is recovered in both versions of our model with the same qualitative results. This is also true for the case of figure \ref{fig:free_m2}(c), where we have decreased the interaction radius from $\Delta = 1$ of figure \ref{fig:free_m2}(a) to 0.8 --- this value is still large enough to keep the system within the small velocity regime. At this point, it is illustrative to see what happens with the steady-state-accumulated normalized distribution $P'(y)$ of the individual particles distance to the centre of mass, defined as:
\begin{equation}
y_i(t) = -\,\hbox{sign}[u_{\rm CM}(t)] \cdot [x_i(t) - x_{\rm CM}(t)].
\label{eq:y_i}
\end{equation}
The results for the different cases are shown in figure \ref{fig:free_m2}(d) for both versions of our model. One can appreciate that the spreading of particles around the centre of mass increases with the noise intensity $\eta$, when comparing the black solid and dotted curves versus the dashed one. Our results show that the width of the distribution $P'(y)$ also increases with the number of particles $N$, while it decreases as the centering parameter $\mu$ becomes larger. In consequence, for the cases with $\Delta=1$ (cases (a) and (b) in figure \ref{fig:free_m2}), as noise increases from $\eta=7$ to $9$, the alternating flock phase is recovered. On the other hand, for the cases with $\eta=7$ (cases (a) and (c) in figure \ref{fig:free_m2}), the width of $P'(y)$ for Version 1 and Version 2 are practically the same (black solid and dotted curves), regardless of the value of $\Delta$. Thus, reducing the value of the latter from 1 to 0.8 allows us to recover the alternating flock phase.

From this analysis, one can conclude that the important quantities to compare, in both versions of our model, is the interaction radius of the alignment interaction $\Delta$ with the width of the distribution $P'(y)$. Moreover, one can also understand why these two versions behave qualitatively in the same way regarding the collective motion of the group, even though Version 1 has an extra source of noise. Looking at the width of the distribution $P'(y)$ for $\eta=0$ and $\Delta=1$, filled solid grey curves in figure \ref{fig:free_m2}(d), one can notice that the width of $P'(y)$ is larger for Version 1 than for Version 2. However, it is much less than the spread induced by the simple additive noise $\xi_i$ in (\ref{eq:Version1}) and (\ref{eq:Version2}) for $\eta>0$. In contrast, the extra source of noise in Version 1, coming from the probabilistic application of alignment and centering interactions, has stronger effects on the fluctuations of the order parameter $\phi(t)$ (see figure \ref{fig:m1e0ss}(d)), i.e., on the values of the individual particles velocities $u_i$ as shown by the widths of the peaks in the distribution $P(u)$ for the cases presented in figure \ref{fig:m1e0ss}(c) and discussed before.

Let us not forget that in all of the models derived from the Czir\'ok \etal model discussed so far (including ours) that present directional switching, the sources of noise are all introduced in the velocity terms of the particles. Moreover, in these models (excluding ours), the ranges of the interactions are all local. What has not been made clear in previous studies is the impact of the different sources of stochasticity on the diffusion of the particles in position space in comparison to the ranges of the different interactions considered. This may be important in relation to how correlations in the motion of particles are kept in time for different systems and the directional switching phenomena. Further studies are being made in this direction.

Additionally, another difference between both versions of our model becomes evident when looking at the order parameter $\phi_{\rm ss}$ in figures \ref{fig:ordpam}(b) and \ref{fig:ordpam}(c), and their corresponding details in figures \ref{fig:ordpam}(e) and \ref{fig:ordpam}(f), as the curves corresponding to Version 2 of our model (curves with clear symbols) are always below to the corresponding ones for Version 1 (curves with solid black symbols), even for different system sizes. This is somewhat counterintuitive in view of the previous results, as the noisier version of our model, Version 1, develops higher mean speeds than Version 2 in general. As explained before, the order parameter $\phi_{\rm ss}$ is always smaller than 1 for both versions of our model, even in the absence of the additive noise ($\eta=0$). A plausible explanation leads us back to the two peaks shown by $P(u)$ in the steady state for this case (see figure \ref{fig:m1e0ss}(c)). While those for Version 1 of our model are symmetric, they are also wider and clearly asymmetric in height, i.e., there is a peak that dominates the distribution centered around a larger value than the narrower and more-balanced-in-height peaks of $P(u)$ for Version 2. Considering that these peaks correspond to the distribution of the velocities of the individual particles, this fact could account for the difference in the order parameter $\phi_{\rm ss}$ for the different cases of the two versions of our model --- in principle, one could obtain the accumulated steady state order parameter as $\phi_{\rm ss} = |\int_{-\infty}^{\infty}u\,P(u)\,du|$.

At this point, it is worth to draw our attention to the cases with $\eta=7$ in the upper plots of figures \ref{fig:v1d-mu05rho1}(d), \ref{fig:v1d-mu1rho1}(d) and \ref{fig:v1d-mu15rho1_v2}(d), where $P(u)$ becomes almost flat and symmetric, very similar to the case without centering ($\mu=0$) shown in the upper plot of figure \ref{fig:v1d-rho1}(d). In the latter, this distribution corresponds to the strong-noise-induced disordered phase, characterized by a random velocity field, where the velocity of every individual particle should average to zero throughout the whole evolution of the system. This is also true in the presence of centering ($\mu>0$) if we were to look just at the distribution $P(u)$. However, there is a big difference when we focus on the collective behaviour of the system. Whereas the mean velocity of the group tends to zero for $\mu=0$, as $\mu$ increases, the mean velocity of the group also increases with the noise amplitude as explained before. One could wonder about the diffusion of the centre of mass for $\mu>0$ and the net displacement it may develop in the long run, or at least, the mean square displacement.

\subsection{Scaling
\label{ssec:scaling}}

Let us now briefly analyze how the properties of both versions of our model scale with the size of the system $N$. First, we focus on the accumulated order parameter $\phi_{\rm ss}$. As shown in figures \ref{fig:ordpam}(b) and \ref{fig:ordpam}(c), and their corresponding details \ref{fig:ordpam}(e) and \ref{fig:ordpam}(f), for weak noise (including $\eta=0$) the curves are rather flat regarding their dependance on $\eta$ and show almost no difference as $N$ changes. This is followed by a crossover region and, as $\eta$ increases, the order parameter increases with $N$. In second place, we have already mentioned, in subsection~\ref{ssec:noise}, that the spread of the position of particles around the centre of mass (the width of the distribution $P'(y)$) increases with $N$ as well. This leads to the conclusion that larger systems will tolerate larger values of $\mu$ before the alternating flock phase is killed, given the interaction radius $\Delta$ of the alignment interaction is kept constant. In third place comes the frequency of the changes in direction in the same phase. As shown in figure \ref{fig:ordpam}(d) for both versions of our model (solid black lines for Version 1 and solid grey lines for Version 2), as $N$ increases, the frequency of the changes in direction becomes smaller. In other words, the intervals between changes in direction increase with the size of the system. This is consistent with the trends reported for some other models that also show directional switching \cite{Yat09, Olo99}.

Overall, both versions of our model scale in the same way. Moreover, aside from the effects of the extra source of noise in Version 1, and fluctuations intrinsic to any numerical analysis, the functional dependence of the quantities mentioned in the previous paragraph on some of the parameters (e.g., the dependence of $\phi_{\rm ss}$ on $\eta$) seem to be quantitatively equivalent across different system sizes. On the other hand, since we only analyzed three system sizes, $N=500, 1000, 4000$, it is no possible for us to provide how these quantities depend explicitly on $N$. We will leave this analysis for a future work, given that all our results point to the fact that it is equivalent to study Version 1 or Version 2 of our model regarding the collective behaviour of the system and its scaling.

\section{Conclusions}
We have successfully introduced a long-range centering interaction in the model of Czir\'ok \etal \cite{Mik99} via two approaches for applying centering and alignment, that derives in two characteristic condensed states. The first, for weak noise, consists in a coherent state with broken symmetry. Here, the fluctuations induced by the combination of centering and alignment in the mean velocity of the densest cluster seem to be stronger than the noise-induced ones. Thus, the system reaches a steady-state speed smaller than the prescribed one even in the absence of noise. The second, for strong noise, consists in an alternating flock phase with non-vanishing mean velocity. In this case, the centering interaction combined with the alignment one provides an ordering mechanism for the densest cluster that effectively performs oscillations around the centre of mass as it changes its direction of motion alternatively. This phase resembles alternating states shown by some other models \cite{Buh06, Yat09, Olo99, Ray06, Bod10}, but also the oscillatory noisy state of the Mikhailov and Zanette model \cite{Mik99}.

The two versions of our model, one where alignment and centering interactions are applied in a probabilistic way and one deterministic, have allowed us to address the directional switching phenomenon that seems to be an intrinsic property of the motion of animal groups as recent experimental results show \cite{Buh06,Yat09}. Even though the origin of the stochasticity in real systems is far from clear, we were able to introduce and understand a new mechanism for directional switching in 1D SPP systems with long-range centering. From our results, new insight has been gained into the effects of different sources of stochasticity on the diffusion of particles in position and velocity spaces, and its relation with the ranges of the different interactions considered for the development of alternating flock phases. In these terms, we have shown how the alternating flock phase of our model can be suppressed and how to recover it.

On the other hand, due to the long-range character of the centering interaction introduced, the properties of our model do not depend on the mean density of particles, but only on the number of particles. In consequence, periodic boundary conditions, typically used in 1D SPP models \cite{Buh06, Yat09, Olo99, Ray06, Czi99}, can be dropped as the system is able to maintain its cohesion. Nonetheless, the scaling properties of our model show the same trends as those reported for some other models \cite{Yat09, Olo99, Bod10}. A more detailed scaling analysis is in the works in order to determine their functional dependence.

Finally, as our model does not show a truly disordered phase (with a random velocity field), it would be worth to study the diffusion of the centre of mass, e.g., its net and mean square displacements throughout the time evolution of the system. We believe our results may be relevant for the understanding of the directional switching phenomenon and, in general, for the theory of flocking.

\ack
This work was partially supported by CONACyT (Mexico), by SEP (Mexico) under grant PROMEP/103.5/10/7296 and by the NSF under grant INT-0336343. We are also grateful for the comments and suggestions from the three anonymous referees.



\section*{References}
\begin{thebibliography}{99}
\bibitem{Lor00} Lorch P D and Gwynne D T 2000 {\it Naturwiss} {\bf 87} 370
\bibitem{Bon98} Bonabeau E, Dagorn L and Fr\'eon P 1998 {\it J. Phys. A: Math. Gen.} {\bf 31} L731
\bibitem{Par99} Parrish J K and Edelstein-Keshet L 1999 {\it Science} {\bf 284} 99
\nonum Parrish J K, Viscido S V and Gr\"unbaum D 2002 {\it Biol. Bull.} {\bf 202} 296
\bibitem{Cou05} Couzin I D, Krause J, Franks N R and Levin S A 2005 {\it Nature} {\bf 433} 513
\bibitem{Buh06} Buhl J, Sumpter D J T, Couzin I D, Hale J J, Despland E, Miller E R and Simpson S J 2006 {\it Science} {\bf 312} 1402
\bibitem{Yat09} Yates C A, Erban R, Escudero C, Couzin I D, Buhl J, Kevrekidis I G, Maini P K and Sumpter D J T 2009 {\it PNAS} {\bf 106} 5464
\bibitem{Wu00} Wu X-L and Libchaber A 2000 \PRL {\bf 84} 3017
\bibitem{Rey87} Reynolds C W 1987 {\it Compurter Graphics} {\bf 21} 4
\bibitem{Ton05} Toner J, Tu Y and Ramaswamy S 2005 {\it Ann. Phys.} {\bf 318} 170
\nonum Vicsek T, Czir\'ok A, Farkas I J and Helbing D 1999 {\it Physica A} {\bf 274} 182
\bibitem{Vic95} Vicsek T, Czir\'ok A, Ben-Jacob E, Cohen I and Shochet O 1995 \PRL {\bf 75} 1226
\bibitem{Gre03} Gr\'egoire G, Chat\'e H and Tu Y 2003 {\it Physica D} {\bf 181} 157
\nonum Gr\'egoire G and Chat\'e H 2004 {\it Phys. Rev. Lett} {\bf 92} 025702
\bibitem{Nag07} Nagy M, Daruka I and Vicsek T 2007 {\it Physica A} {\bf 373} 445
\bibitem{Per08} Peruani F and Sibona G J 2008 \PRL {\bf 100} 168103 
\bibitem{Bus97} Bussemaker H J, Deutsch A and Geigant E 1997 \PRL {\bf 78} 5018
\bibitem{Ald03} Aldana M and Huepe C 2003 {\it J. Stat. Phys.} {\bf 112} 135
\nonum Aldana M, Dossetti V, Huepe C, Kenkre V M and Larralde H 2007 \PRL {\bf 98} 095702
\bibitem{Olo99} O'Loan O J and Evans M R 1999 {\it J. Phys. A: Math. Gen.} {\bf 32} L99
\bibitem{Ray06} Raymond J R and Evans M R 2006 {\it Phys. Rev. E} {\bf 76} 036112
\bibitem{Bod10} Bode N W F,  Franks D W and Wood A J 2010 {\it J. theor. Biol.} {\bf 267} 292
\bibitem{Mik99} Mikhailov A S and Zanette D H 1999 {\it Phys. Rev. E} {\bf 60} 4571
\bibitem{Erd05} Erdmann U, Ebeling W and Mikhailov 2005 {\it Phys. Rev. E} {\bf 71} 051904
\bibitem{Shi96} Shimoyama N, Sugawara K, Mizuguchi T, Hayakawa Y and Sano M 1996 \PRL {\bf 76} 3870
\bibitem{Lev00} Levine H, Rappel W-J and Cohen I 2000 {\it Phys. Rev. E} {\bf 63} 017101
\bibitem{Dor06} D'Orsogna M R, Chuang Y L, Bertozzi A L and Chayes L S 2006 \PRL {\bf 96} 104302
\bibitem{Dos09} Dossetti V, Sevilla F J and Kenkre V M 2009 {\it Phys. Rev. E} {\bf 79} 051115
\bibitem{Str08} Strefler J, Erdmann U and Schimansky-Geier L 2008 {\it Phys. Rev. E} {\bf 78} 031927
\bibitem{Iwa10} Iwasa M, Iida K and Tanaka D 2010 {\it Phys. Rev. E} {\bf 81} 046220
\bibitem{Top04} Topaz C M and Bertozzi A L 2004 {\it SIAM J. Appl. Math.} {\bf 65} 152
\bibitem{Ber06} Bertin E, Droz M and Gr\'egoire G 2006 {\it Phys. Rev. E} {\bf 74} 022101
\nonum Bertin E, Droz M and Gr\'egoire G 2009 {\it J. Phys. A: Math. Theor.} {\bf 42} 445001
\bibitem{Mog96} Mogilner A and Edelstein-Keshet L 1996 {\it Physica D} {\bf 89} 346
\bibitem{Gon08} G\"onci B, Nagy M and Vicsek T 2008 {\it Eur. Phys. J. Special Topics} {\bf 157} 53
\bibitem{Czi99} Czir\'ok A, Barab‡si A-L, and T. Vicsek 1999 \PRL {\bf 82} 209


\end{thebibliography}
\end{document}